\documentclass[12pt,manuscript]{aastex}
\usepackage{apjfonts}

\newcommand{\msun}{M_\odot}

\newcommand\ha{H$\alpha$}
\newcommand\sii{[S\thinspace {\footnotesize II]}}
\newcommand\oiii{[O\thinspace {\footnotesize III]}}

\newcommand\lam{\mbox{$\:\lambda $ }}
\newcommand\lamlam{\mbox{$\:\lambda\lambda $ }}

\newcommand{\gapprox}{\mathrel{\mathpalette\@versim>}}
\newcommand{\lapprox}{\mathrel{\mathpalette\@versim<}}
\newcommand{\propapprox}{\mathrel{\mathpalette\@versim\propto}}

\shorttitle{N103B: A Type Ia with CSM}
\shortauthors{WILLIAMS ET AL.}

\begin{document}

\title{Spitzer Observations of the Type Ia Supernova Remnant N103B:
  Kepler's Older Cousin?}

\author{Brian J. Williams,\altaffilmark{1,2}
Kazimierz J. Borkowski,\altaffilmark{3}
Stephen P. Reynolds,\altaffilmark{3}
Parviz Ghavamian,\altaffilmark{4}
John C. Raymond,\altaffilmark{5}
Knox S. Long,\altaffilmark{6}
William P. Blair,\altaffilmark{7}
Ravi Sankrit,\altaffilmark{8}
P. Frank Winkler\altaffilmark{9}
\& Sean P. Hendrick\altaffilmark{10}
}

\altaffiltext{1}{NASA Goddard Space Flight Center, Greenbelt, MD 20771; brian.j.williams@nasa.gov}
\altaffiltext{2}{NASA Postdoctoral Program Fellow}
\altaffiltext{3}{Physics Dept., North Carolina State University,
    Raleigh, NC 27695-8202.}
\altaffiltext{4}{Dept. of Physics, Chemistry, and Geosciences, Towson University, Towson, MD 21252}
\altaffiltext{5}{Harvard-Smithsonian Center for Astrophysics, 60 Garden
    Street, Cambridge, MA 02138.}
\altaffiltext{6}{STScI, 3700 San Martin Dr., Baltimore, MD 21218}
\altaffiltext{7}{Dept. of Physics and Astronomy, Johns Hopkins University, 
    3400 N. Charles St., Baltimore, MD 21218-2686}
\altaffiltext{8}{SOFIA/USRA}
\altaffiltext{9}{Dept. of Physics, Middlebury College, Middlebury, VT 
    05753}
\altaffiltext{10}{Physics Dept., Millersville U., PO Box 1002, Millersville, 
    PA 17551}

\begin{abstract}

We report results from {\it Spitzer} observations of SNR 0509-68.7,
also known as N103B, a young Type Ia supernova remnant in the Large
Magellanic Cloud that shows interaction with a dense medium in its
western hemisphere. Our images show that N103B has strong IR emission
from warm dust in the post-shock environment. The post-shock gas
density we derive, 45 cm$^{-3}$, is much higher than in other Type Ia
remnants in the LMC, though a lack of spatial resolution may bias
measurements towards regions of higher than average density. This
density is similar to that in Kepler's SNR, a Type Ia interacting with
a circumstellar medium. Optical images show H$\alpha$ emission along
the entire periphery of the western portion of the shock, with [O III]
and [S II] lines emitted from a few dense clumps of material where the
shock has become radiative. The dust is silicate in nature, though
standard silicate dust models fail to reproduce the ``18 $\mu$m''
silicate feature that peaks instead at 17.3 $\mu$m. We propose that
the dense material is circumstellar material lost from the progenitor
system, as with Kepler. If the CSM interpretation is correct, this
remnant would become the second member, along with Kepler, of a class
of Type Ia remnants characterized by interaction with a dense CSM
hundreds of years post-explosion. A lack of N enhancement eliminates
symbiotic AGB progenitors. The white dwarf companion must have been
relatively unevolved at the time of the explosion.


\end{abstract}

\section{Introduction}
\label{intro}

While the importance of Type Ia supernovae (SNe) in astrophysics is
widely known, there are significant uncertainties in the nature of the
progenitor systems and the explosions themselves. There is now
significant debate in the literature over whether these SNe result
from single-degenerate (the explosion of a white dwarf that has
accreted matter close to the Chandrasekhar mass of 1.4 $\msun$ from a
companion) or double-degenerate (the merger of two sub-Chandrasekhar
mass white dwarfs) systems. In the last few years, most studies have
favored a double-degenerate origin for most SNe Ia; \citet{gilfanov10}
claim that 95\% of SNe Ia must result from this progenitor
system. \citet{wang12} provide a thorough review of the possible
explosion mechanisms for Type Ia SNe.

Recently, particular attention has been paid to a growing subclass of
extragalactic Type Ia SNe that show signs of interaction with material
in a circumstellar medium (CSM) at early times
\citep{silverman13}. Although only about a dozen of these objects have
been observed, the host galaxies are all late-type spirals like the
Milky Way or dwarf irregulars like the Large Magellanic Cloud (LMC),
implying an origin in relatively young stellar populations
\citep{silverman13}. Observations of these SNe imply that the
surrounding CSM is quite dense; however, to date, no radio or X-ray
emission has been observed (likely due to their being too far
away). Supernova remnants (SNRs), on the other hand, may be able to
manifest the presence of CSM hundreds or even thousands of years after
the explosion.

We report here on Spitzer imaging and spectroscopic observations of
N103B, a small ($\sim 15''$ radius, or $\sim 3.6$ pc at a distance of
50 kpc) SNR in the LMC. Light echoes from the remnant place its age at
$\sim 870$ years \citep{rest05}, or about twice as old as Kepler's
SNR. As we show in this paper, this remnant bears some strong
resemblances to Kepler's SNR in our Galaxy, so much so that we propose
that it may be a Kepler analog at 50 kpc.

Also known as SNR 0509-68.7, N103B is not without its share of
controversy. Arguments for a Type Ia origin have been made by several
authors. \citet{hughes95} examined {\it ASCA} spectra and found a high
Fe/O ratio in the X-ray spectra. \citet{lewis03} verified this result
with {\it Chandra} observations. \citet{lopez11} conducted a
statistical analysis of the morphology of the remnant and concluded
that the overall shape of the emission was more consistent with that
of Type Ia SNRs in both the LMC and the Galaxy. \citet{yang13} favor a
Type Ia origin based on the strength of Cr and Mn lines in the
spectrum and their relative strength compared to Fe. Additionally,
\citet{badenes09} favor a Type Ia SNR, noting that the remnant is
associated with a region of the LMC that underwent recent star
formation, implying that the progenitor, like Kepler, might have had a
relatively young, massive progenitor with substantial mass loss prior
to explosion. They suggest that the structure of the remnant may be
due to CSM interaction. Finally, no compact remnant or pulsar wind
nebula has ever been observed in the remnant, whereas many known
core-collapse (CC) SNRs in the LMC do have known neutron stars or
pulsars (a notable exception to this is SN 1987A, where no compact
remnant has yet been observed).

On the other hand, \citet{vanderheyden02} argue in favor of a CC
supernova origin for N103B based on the presence of O lines in the
{\it XMM-Newton} RGS spectra. Most recently, \citet{someya13} find
that the {\it Suzaku} X-ray spectra are best fit with H-dominated
plasmas that more likely originate in Type II SNe. They find that the
abundances in this model are most consistent with a 13 $\msun$
progenitor.

However, recently obtained light echo spectroscopy has settled the
issue. \citet{rest14} show that the spectra of the light echos from
N103B, first reported in \citet{rest05}, are consistent with a SN
Ia. {\em Light echo confirmation of the SN type is something that not
  even Kepler has.}

From the SNR perspective, remnants of Type Ia SNe can generally be
modeled without invoking a dense or complex CSM in their surroundings
\citep{badenes07}. Most Type Ia remnants, such as SN 1006, are
evolving into a rather low-density ISM
\citep{katsuda13,winkler13}. Other examples of such remnants include
Tycho's SNR \citep{williams13}, SNR 0509-67.5 \citep{williams11}, and
SN 1885 in M31 \citep{badenes07}. Searches for potential companion
stars to these SNe have come up empty
\citep[e.g.,][]{schaefer12,kerzendorf14}. The simplest explanation for
these (and indeed, most) Type Ia remnants is the double-degenerate
scenario described above.

There are a few outliers: RCW 86 appears to have exploded into a low
density, wind-blown cavity \citep{williams11b}, implying an
single-degenerate scenario. Kepler's SNR, on the other hand, has
extremely dense surroudings, indicative of CSM
material. \citet{blair07} and \citet{williams12} used {\it Spitzer}
imaging and spectroscopic observations to quantify the high densities
in Kepler. From the spectral signatures of the dust observed, they
concluded that an O-rich AGB star must have been part of the
progenitor system. \citet{reynolds07} and \citet{burkey13} concluded
from {\it Chandra} X-ray observations that Kepler was an example of a
``prompt'' Type Ia SN; i.e., a progenitor that exploded within a few
hundred Myr of forming, and that the remnant is most consistent with a
progenitor system which blew a dense, equatorial wind prior to
exploding, a situation which implies a single-degenerate progenitor
scenario. Thus far, Kepler's SNR is the only remnant known with a
dense CSM; finding others will bridge the gap between SNe Ia-CSM and
their remnants.

We organize this paper in the following fashion. In
Section~\ref{observations}, we detail the infrared and optical
observations and data reduction. In Section~\ref{results}, we report
the results of our analysis. Section~\ref{disc} provides a discussion
of our interpretation of our results. Section~\ref{conclusions} serves
as a summary of our findings.

\section{Observations}
\label{observations}

{\it Spitzer} imaging observations of N103B were carried out as part
of Program ID 3680. We observed the remnant in all bands of both the
Infrared Array Camera (IRAC) and the Multiband Imaging Photometer for
Spitzer (MIPS). IRAC images were obtained on 2004 Dec 19, with MIPS
imaging occuring on 2005 Apr 8. Results for N103B from this observing
campaign were first published as part of an infrared SNR atlas of the
LMC by \citet{seok13}, who report a flux at 24 $\mu$m of 0.505 Jy. The
remnant is small enough ($30''$ in diameter) to easily fit within the
FOV of all of {\it Spitzer's} instruments. The spatial resolution of
{\it Spitzer} is $1'' - 2''$ for the IRAC arrays, $\sim 7''$ for the
24 $\mu$m camera, and $20''$ at 70 $\mu$m. Our IRAC observations
consisted of a dither pattern of five pointings with a frame time of
30 seconds for each frame. MIPS observations depended on the camera
used. At 24 $\mu$m, we mapped the remnant with 44 overlapping
pointings of 10 s each. At 70 $\mu$m, we mapped it with 94 pointings
of 10 s each. The 160 $\mu$m map consisted of 252 pointings of 3 s. As
the 160 $\mu$m images show only emission from the ISM with no
discernable emission from the remnant, we do not discuss them
further. Likewise, no obvious emission is seen at wavelengths at 100
$\mu$m or beyond in archival {\it Herschel} data. All {\it Spitzer}
data have been processed with the {\it Spitzer} Post-Basic Calibrated
Data pipeline, version 19.0.

We used the Image Co-addition with Optional Resolution Enhancement
(ICORE) software to deconvolve the 24 $\mu$m image. ICORE is publicly
available via the web\footnotemark[11], and is designed for infrared
astronomy. We processed our 24 $\mu$m data with the software, which
uses the known PSF of {\it Spitzer} and all the individual frames that
go into making a mosaic to deconvolve the image. Because we had such
good coverage of the remnant with our 44 overlapping frames, we were
able to obtain a resolution of $\sim 2''$ (based on measuring the FWHM
of several point sources in the field). This allows for a more
detailed comparison with X-ray and optical images of the remnant. 

\footnotetext[11]{http://web.ipac.caltech.edu/staff/fmasci/home/icore.html}

Spectroscopic observations with the Infrared Spectrograph (IRS) were
obtained on 2008 Apr 26 as part of Program ID 40604. We mapped the
remnant using the long-wavelength (14-38 $\mu$m), low-resolution
($\delta \lambda/\lambda$ 64-128) (LL) spectrograph. We stepped across
the remnant in eight LL slit pointings, stepping perpendicularly
5.1$''$ each time. This step size is half the slit width and is also
the size of a pixel on the LL instrument. We then shifted the slit
positions by 56$''$ in the parallel direction and repeated the map,
ensuring sufficient redundancy for each spatial location. This process
was repeated for each of the two orders of the spectrograph. For the
shorter wavelength order (14-21 $\mu$m), each pointing consisted of
two cycles of 30 s integration time. The longer wavelength order
(where the remnant is brighter) consisted of two cycles of 14 s
integration times. We also obtained short-wavelength (5.2-14.5
$\mu$m), low-resolution (SL) spectroscopy of a few select locations in
the remnant. The choice of these locations was guided by our imaging
observations from 2005. We use the IRS contributed software {\it
  CUBISM}\footnotemark[12] \citep{smith07} to extract 1-D spectra from
both the source and the background. For the spectra shown here, we
subtract off a global background obtained by extracting simultaneous
off-source spectra from either side of the remnant and averaging them
together.

\footnotetext[12]{http://irsa.ipac.caltech.edu/data/SPITZER/docs/dataanalysistools/tools/cubism/}

Additionally, we used the IRS Peak-up array at 16 $\mu$m to obtain an
image of the remnant at $\sim 4''$ resolution. The Peak-up mapping
consisted of a dithered map of five pointings, where each pointing was
made up of four 30 s integration images. Because we already had an
image at 24 $\mu$m, we did not use the 22 $\mu$m Peak-up camera. All
{\it Spitzer} MIPS and Peak-up images are shown in
Figure~\ref{images}.

\subsection{Optical Observations}
\label{opticalobs}

Optical observations of N103B were carried out from the 1.5 m
telescope at CTIO in 1994.  These included narrow-band images in \ha,
\sii\lamlam 6716,6731, and \oiii\lam 5007, plus red and green
continuum filters for subtracting the stars, as detailed in
Table~\ref{optical}. The images were processed, including bias
subtraction, flat-fielding, combining, and re-projection onto a common
world coordinate system using standard IRAF\footnotemark[13]
techniques.  The continuum frames were then subtracted from the
emission-line ones (green from \oiii, and red from \ha\ and \sii), and
the subtracted images were flux-calibrated using standard stars from
\citet{hamuy92}.  The seeing for all the observations was about
1\farcs 6.

\footnotetext[13]{IRAF is distributed by the National Optical
  Astronomy Observatories, which is operated by the AURA, Inc. under
  cooperative agreement with the National Science Foundation.}

The continuum-subtracted image (in all three lines, displayed in
colors indicated in the caption), is shown in Figure~\ref{images2}.
The filament shown in red that wraps around the entire western portion
of the shell is quite faint and appears only in \ha, the likely
signature of a nonradiative shock, while the much brighter knots that
appear yellow are almost equally strong in \sii\ and \ha, indicating
that here the shocks have become radiative in much denser material.
The very bright white knot (indicated in Figure~\ref{images2}) is
strong in all three lines, as well as in the infrared.

\section{Results and Analysis}
\label{results}

\begin{figure}
\includegraphics[width=15cm]{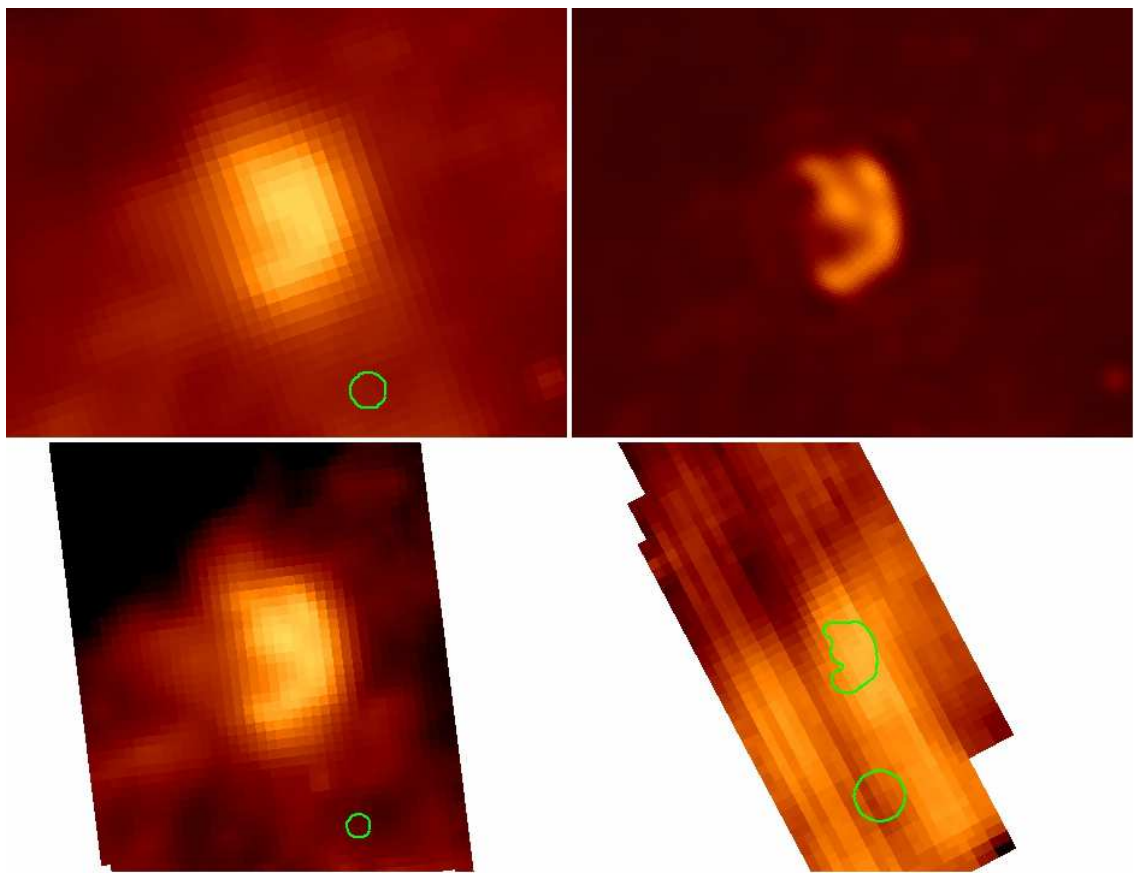}
\caption{{\it Spitzer} images of N103B. {\it Top Left}: 24 $\mu$m
  image; {\it Top Right}: 24 $\mu$m image, deconvolved (see text for
  details); {\it Bottom Left}: 16 $\mu$m ``Peakup'' array image; {\it
    Bottom Right}: 70 $\mu$m image, zoomed out by a factor of two to
  show the surroundings of the remnant. A single 24 $\mu$m contour is
  overlaid. The native resolutions of all three instruments are shown
  as circles on their respective images. For all images here and in
  subsequent figures, north is up and east is to the left.
\label{images}
}
\end{figure}

In Figure~\ref{images}, we show the results of our IR imaging
observations. From a circular aperture $40''$ in diameter, we measure
a 24 $\mu$m flux of 0.48 Jy. By experimenting with several different
background apertures, we can get a measure of the statistical error on
the flux, which we find to be only 2\%. The remnant is extremely
bright at 24 $\mu$m, particularly compared to the surrounding ISM,
which itself is quite uniform. The {\it MIPS Instrument Handbook}
lists calibration uncertainties on fluxes of extended sources at 4\%,
so our overall uncertainty should be $\lesssim 0.03$ Jy. The [O
  IV]/[Fe II] line complex at 26 $\mu$m contributes a negligible flux
when the spectrum is integrated over the wide MIPS 24 $\mu$m bandpass;
i.e., virtually all of the emission seen is from dust. Our flux
measurement is within errors of that from \citet{seok13}. We list all
fluxes in Table~\ref{irflux}.

At 16 $\mu$m, the remnant is still quite bright. We measure a flux of
0.19 Jy, with statistical uncertainties of 4\%. The {\it IRS Data
  Handbook} reports systematic uncertainties of 6\% on measured fluxes
from the Peak-Up array, so combined in quadrature with our measurement
uncertainties, we find an overall uncertainty in the 16 $\mu$m flux of
0.014 Jy. For both the 24 and 16 $\mu$m fluxes, we have used the
images from the {\it Wide-Field Infrared Survey Explorer} (WISE)
archive as a cross-calibration. At 22 and 12 $\mu$m, we measure fluxes
of 0.41 and 0.13 Jy, respectively.

At 70 $\mu$m, the situation becomes more complicated. The PSF of the
telescope is less than the diameter of the remnant, but only
barely. However, the background surounding N103B emission is quite
complex at 70 $\mu$m, as is shown in Figure~\ref{images}. Throughout
the MIPS 70 $\mu$m field of view, the surface brightness levels in the
image vary from $\sim 30 - 130$ MJy/sr. At the location of the
remnant, there is a peak in the emission, but other peaks of
comparable brightness exist that are unrelated to the remnant, most
notably to the E and SW. Using aperture photometry with varying
background regions, we obtain fluxes ranging from 0.04 to 1.3
Jy. Given this extremely large spread, we conclude that while there
likely is very faint emission from the remnant at this wavelength, the
flux measurement itself is not reliable enough to be used as a
constraint in modeling the dust in the remnant. We consider the
highest value measured, 1.3 Jy, to be an upper limit to the flux at 70
$\mu$m.

\begin{figure}
\includegraphics[width=16cm]{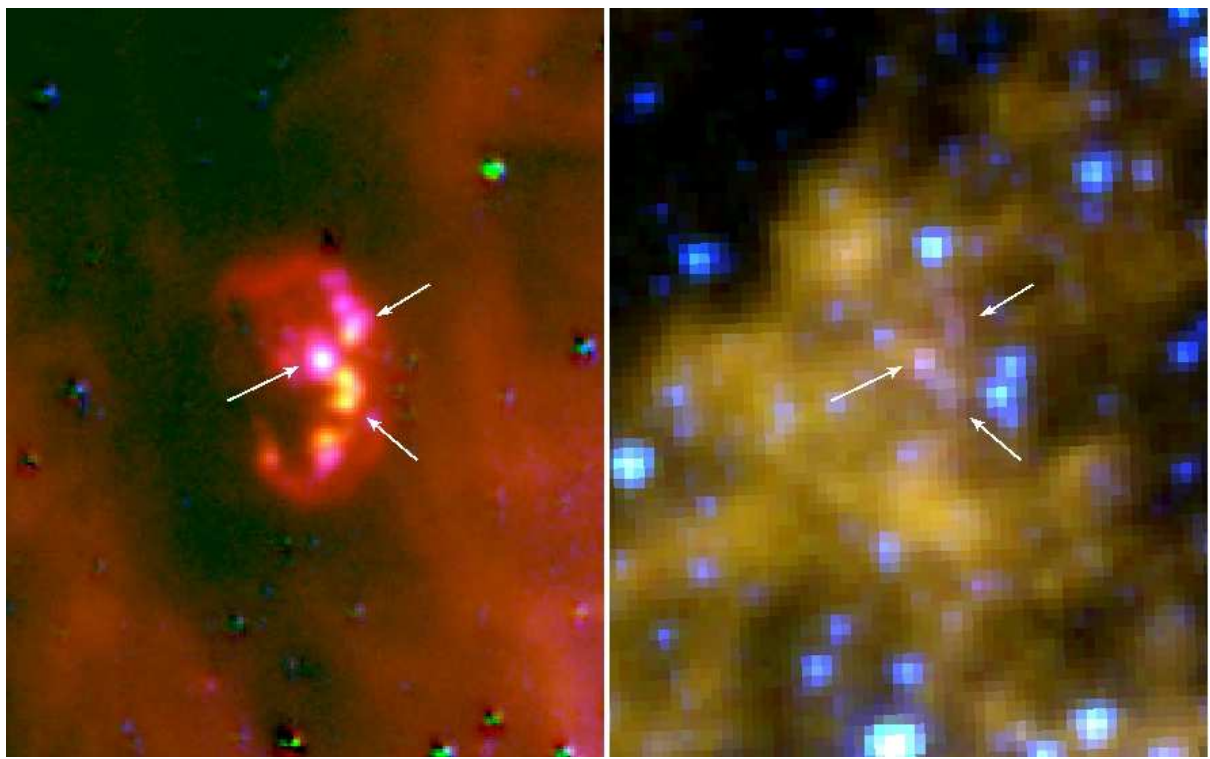}
\caption{{\it Left}: Optical emission-line image of N103B, where red =
  \ha, green = \sii, and blue = \oiii, in a display scaled as the
  square root of the intensity.  The stars have been largely removed
  by subtracting scaled continuum images. {\it Right}: Three-color
  IRAC image, with 8.0 $\mu$m emission in red, 5.6 $\mu$m in green,
  and 4.5 $\mu$m in blue. The colors allow a distinction between
  emission (indicated with arrows) that is likely from the remnant and
  that of the general diffuse ISM. Both panels are on the same scale.
\label{images2}
}
\end{figure}

In Figure~\ref{images2}, we show the results of our IRAC imaging. In
the individual IRAC bands, there is no obvious emission from the
remnant. However, a three-color image, with the 8.0 $\mu$m, 5.6
$\mu$m, and 4.5 $\mu$m images shown in red, green, and blue,
respectively, does bring out emission that not only has slightly
different colors than the ambient ISM, but also corresponds
morphologically with the optical emission from the radiative shocks,
as traced by [S II], also shown in Figure~\ref{images2}.

\begin{figure}
\includegraphics[width=15cm]{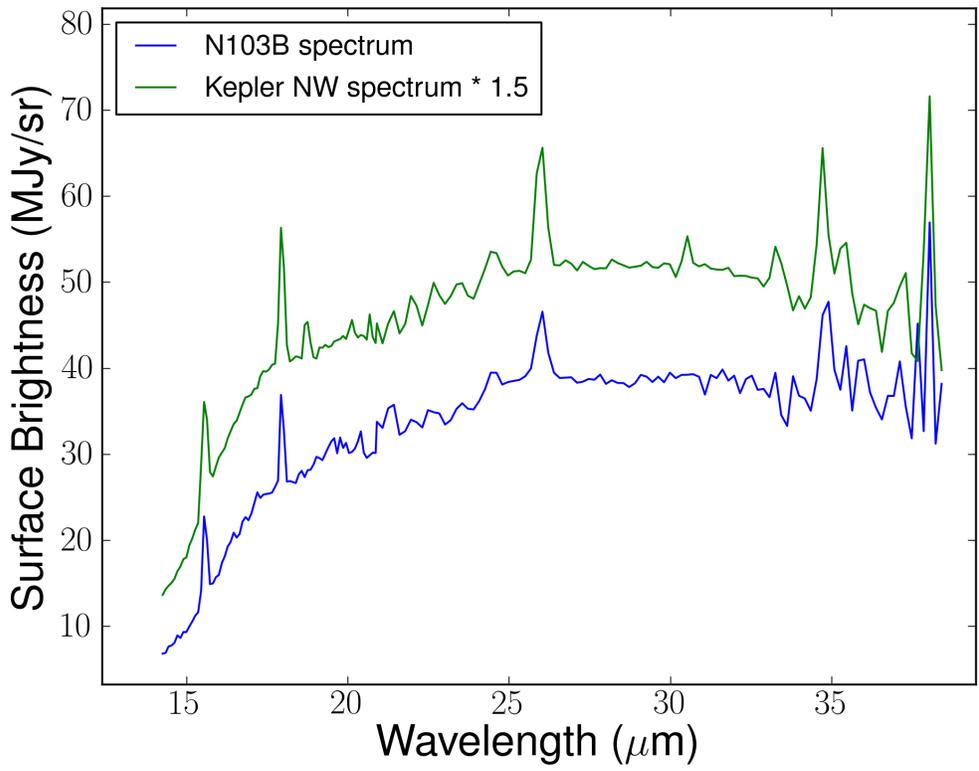}
\caption{IRS LL spectra from N103B and the NW region of Kepler's
  SNR. For display purposes, the spectrum from Kepler has been
  multiplied by a factor of 1.5.
\label{spectra}
}
\end{figure}

In Figure~\ref{spectra}, we show the spatially integrated LL spectrum
of N103B. The spectrum is clearly dominated by continuum emission from
warm dust, and in many ways, looks similar to the spectra from two
other young remnants of Type Ia SNe in the LMC: 0509-67.5 and
0519-69.0 \citep{williams11}. However, unlike these two remnants,
N103B shows line emission from low ionization states of various
elements. [Fe II] lines at 18.7 and 24.3 $\mu$m are identifiable, and
it is likely that most of the line at 26 $\mu$m is due to [Fe II] as
well (the 26 $\mu$m Fe line blends with an [O IV] line at 25.9 $\mu$m
at this spectral resolution, and we do not have high-resolution IR
spectra to separate them). Lines from [Ne III] at 15.5 $\mu$m and [Si
  II] at 34.8 $\mu$m are also easily identifiable in the spectrum; the
spike at $\sim 38$ $\mu$m is an instrumental artifact. It is not
surprising to see such lines in the spectrum of N103B (and not
0509-67.5 or 0519-69.0), because N103B is known to have some radiative
shocks (see Figure~\ref{images2}), whereas the other two young
remnants are still purely in a nonradiative phase
\citep{tuohy82,smith91}. For comparison, we also show the spectrum
from the NW region of Kepler's SNR \citep{williams12}. The
similarities between the two spectra are obvious. Both show emission
from silicate dust, as evidenced by the broad feature at 18 $\mu$m
which manifests itself as a ``shoulder'' in the spectrum.

\subsection{Dust Modeling}
\label{modeling}

In \citet{williams12}, we showed that heating of dust grains in Kepler
by photons from the radiative shocks is insufficient to heat grains to
the emitting temperatures observed. Even with unphysical assumptions,
such as 100\% of the shock energy being converted into UV photons,
heating by this mechanism can only heat grains to at most 50 K (dust
temperatures in N103B, as in Kepler, are much higher than this at
around $\sim 115-130$ K). Collisional heating is required for such hot
dust. Collisional heating in the slow, radiative shocks is
insufficient to heat grains to the high observed temperatures required
to emit in the mid-IR, as shown in \citet{dwek87}. Only the plasma
conditions of the nonradiative, X-ray emitting shocks, traced in the
optical by H$\alpha$ emission, provide an environment for grains to be
heated to such temperatures. This implies a morphological correlation
between these different energy regimes. We do observe such a
correlation; see Section~\ref{multi} for more discussion on this.

To model the dust emission observed, we use the models described by
\citet{borkowski06} and follow the procedures used by
\citet{williams11} and \citet{williams12}. Briefly, grains are heated
via collisions with hot ions and electrons in the post-shock plasma
(the same hot plasma that emits thermal X-rays), where the final grain
temperatures are a function of both the plasma temperature and density
(the stronger dependence). We include effects from the gradual erosion
of grains due to sputtering. The type of dust is a parameter as well,
and due to the fact that we clearly see the 18 $\mu$m silicate dust
emission feature, we have opted to use standard ``astronomical
silicates,'' with optical constants taken from \citet{draine84}. As in
Kepler, we see no evidence for crystalline silicate features, such as
those seen in the spectra around cool, evolved O-rich stars
\citep{morris08,henning10}.

The X-ray plasma parameters are taken from \citet{lewis03}, who list
average values of the electron temperature and ionization timescale of
the plasma in N103B of 1 keV and 10$^{11}$ cm$^{-3}$ s,
respectively. Though we do not have a direct measure of the ion
temperature, we use an approximate value of 5 keV. This is slightly
lower than the temperature we used in Kepler (8.9 keV), but given that
N103B is slightly more evolved than Kepler, it is reasonable to assume
the ion temperature will be lower. We stress here that the plasma
temperature is the least sensitive parameter in collisional heating;
see \citet{williams13} for a fuller discussion of the weak dependence
of the dust spectrum on both the ion and electron temperature. If we
had used the same proton temperature as in Kepler (almost a factor of
two higher), our derived density would be lower by only 20\%.

\begin{figure}
\includegraphics[width=15cm]{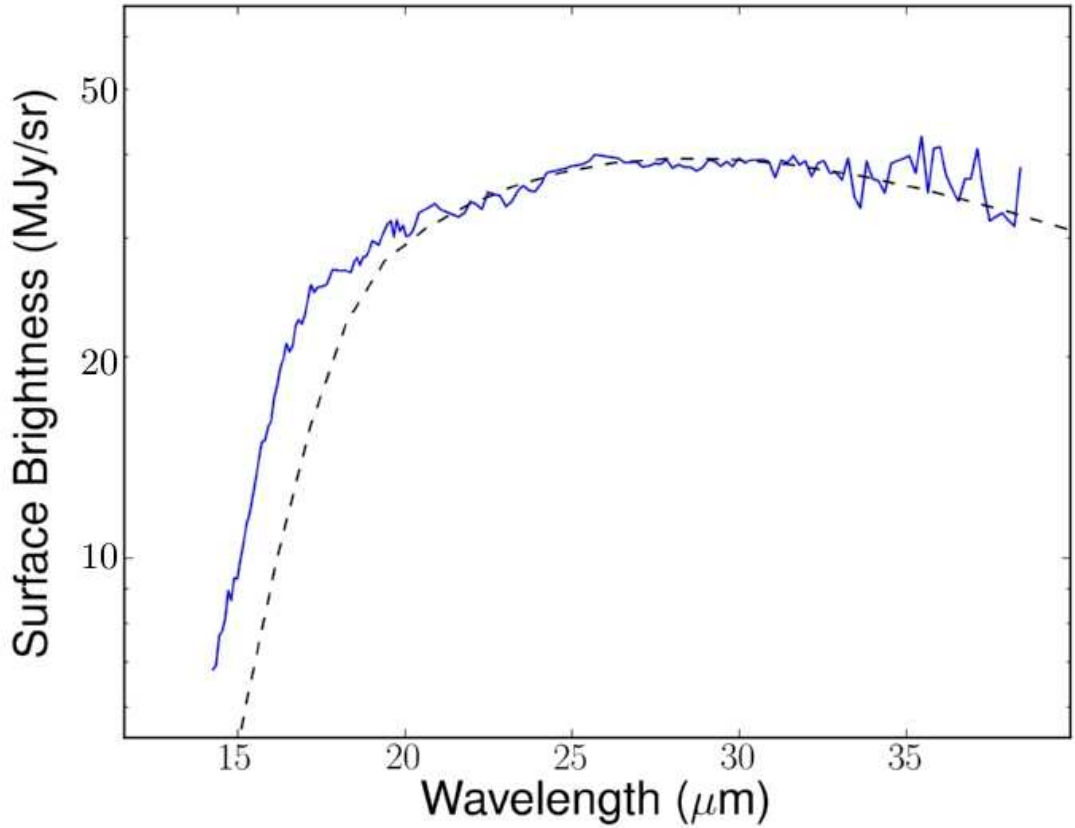}
\caption{LL spectrum of N103B, overlaid with model fit with post-shock
  density, n$_{H}$ = 45 cm$^{-3}$. The spectrum is the same as that
  shown in Figure~\ref{spectra}, except that the lines and bad pixels
  have been removed (to highlight the shape of the continuum) and the
  vertical axis is logarithmic. The model, which includes the 18
  $\mu$m silicate feature, fails to reproduce the observed spectrum
  because the feature in N103B is not located at 18 $\mu$m.
\label{spectralfit}
}
\end{figure}

We use a $\chi^{2}$ minimization algorithm to fit the spectrum from 21
to 33 $\mu$m, which we have found to be the most reliable region of
the spectrum to fit to obtain a gas density. We find a best-fit with a
post-shock density of n$_{H}$ = 45 cm$^{-3}$. The reduced $\chi^{2}$
value for this fit is close to 1 over the fitted region. We obtain a
90\% confidence interval for our upper and lower limits (where $\Delta
\chi^{2}\ = 2.71$) of 40 and 49 cm$^{-3}$, respectively. We show the
fit to the spectrum in Figure~\ref{spectralfit}. This density is quite
high, particularly compared to other Type Ia SNRs in the LMC
\citep{borkowski06} that are believed to be expanding into the ambient
interstellar medium (ISM). We discuss possible explanations for this
in Section~\ref{disc}. In this model, 51\% of the dust is destroyed
via sputtering, the process of gradually liberating material from the
surface of the grain via collisions with energetic ions
\citep{draine79}. We require a current dust mass of 1.5
$\times\ 10^{-3}\ \msun$. Accounting for the amount lost due to grain
destruction, the total swept up dust mass in the remnant is 3
$\times\ 10^{-3}\ \msun$. For a standard LMC dust-to-gas ratio of 2.5
$\times 10^{-3}$ \citep{weingartner01}, this leads to a swept-up gas
mass of 1.2 $\msun$. However, previous studies
\citep{borkowski06,williams06} have shown that the dust-to-gas ratio
as determined from mid-IR observations of SNRs can be lower than the
canonical value by a factor of a few, so the actual swept-up gas mass
in N103B is likely a few solar masses.

\subsection{18 $\mu$m Silicate Feature}

\begin{figure}
\includegraphics[width=15cm]{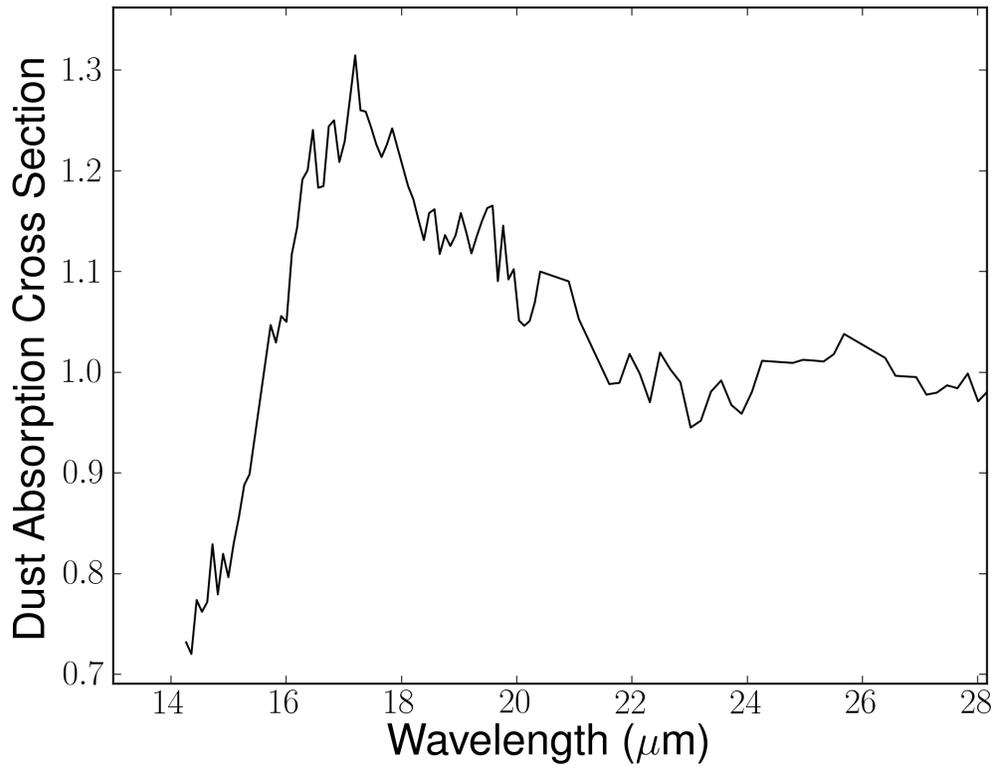}
\caption{Absorption cross section of the dust in N103B obtained by
  dividing spectrum by blackbody fit, as described in the text. The
  broad feature peaks at 17.3 $\mu$m, rather than 18 $\mu$m like most
  interstellar silicate dust.
\label{silicatefeature}
}
\end{figure}

As in Kepler, our dust models have no trouble fitting the continuum
beyond 20 $\mu$m, but between 15 and 20 $\mu$m, the model
underpredicts the observed emission. This is the region of the
spectrum where the ``18 $\mu$m'' silicate feature is observed
\citep{draine84}. Despite the fact that our silicate grain model does
include this feature, it still fails to reproduce the observed
spectrum in this region because the ``18 $\mu$m'' feature is not, in
fact, at 18 $\mu$m. Again, as is the case with Kepler, the silicate
feature is located at approximately 17.3 $\mu$m. We find this by
following the technique of \citet{guhaniyogi11} of fitting a blackbody
model to the underlying continuum and dividing this out of the
spectrum. What remains is the absorption efficiency of the silicate
dust grains as a function of wavelength, which we show in
Figure~\ref{silicatefeature}.

There is no dust temperature at which the optical constants of
\citet{draine84} can account for the observed features. There are
examples in the literature of variation in the location of the
silicate feature. \citet{ossenkopf92} note that there are differences
between silicates in the CSM and ISM, since circumstellar dust is
modified by processing within the ISM. This suggests the possibility
that the dust we are seeing would be newly-formed dust from the
outflow from the progenitor that has not yet been reprocessed within
the ambient ISM. \citet{henning10} report variations in silicate
features seen from the spectra of AGB stellar outflows. Additionally,
\citet{smith10} find that the ``18 $\mu$m'' feature in the nucleus of
M81 is actually at 17.2 $\mu$m, similar to our observed location.

\subsection{X-ray Modeling}

The presence of O lines in the {\it XMM-Newton} RGS grating spectrum
is reported by \citet{vanderheyden02}. These authors inferred from
this that the progenitor must have been a core-collapse SN from a
massive star. They report a value of 0.27 $\msun$ of O assuming a
density that is a few times higher than that which we report
here. With the density in the post-shock environment now determined
from our IR observations reported, we can easily reconcile the
presence of O lines with the fact that N103B is now known to be a SN
Ia. The higher densities we find lower the mass of O implied by the
fits of \citet{vanderheyden02}. It is beyond the scope of this paper
to do our own detailed spectral fitting to the X-ray data, but we note
that it is also possible that some of the O lines come from the
ejecta; e.g., the canonical ``W7'' model of \citet{nomoto84} produces
0.14 $\msun$ of O. However, \citet{lewis03} use the high-resolution
{\it Chandra} data to conclude that it is unlikely that the O
component of the spectrum originates within the ejecta. The bulk of
the O emission seen likely comes from the interaction of the forward
shock with the dense circumstellar material. X-ray O emission arises
quite easily in shocked gas, and even the remnant of SN 1006 (widely
known to be the remnant of a Type Ia SN) shows substantial O lines in
its X-ray spectrum \citep{winkler14}, which may arise from ejecta or
shocked ISM.

Detailed X-ray spectral modeling is beyond the scope of this paper,
but we have fit models to the integrated RGS X-ray spectrum to
determine the shocked gas mass in the remnant. We fit both a
plane-shock model and a Sedov model. For an assumed absorption column
of 2.5 $\times 10^{21}$ cm$^{-2}$, we derive emission measures of 1.3
and 1.5 $\times 10^{59}$ cm$^{-3}$ for the plane-shock and Sedov
models, respectively. With our measured post-shock density of 45
cm$^{-3}$, we derive a total swept-up mass of 2.7-3.3 $\msun$. While
only a rough estimate, this number agrees relatively well with our
inferred gas mass from the assumed dust-to-gas mass ratio.

\subsection{Multi-Wavelength Comparison}
\label{multi}

In Figure~\ref{images2}, we show a three-color optical image of N103B
at $\sim 1.8''$ resolution. Only the western hemiphere of the remnant
is visible, with the entire forward shock emitting in H$\alpha$. As we
mentioned in Section~\ref{opticalobs}, this is strongly suggestive
that the Balmer emission observed in N103B arises from nonradiative
shocks over a significant portion of the remnant.

\begin{figure*}
\centering
\includegraphics[width=12cm]{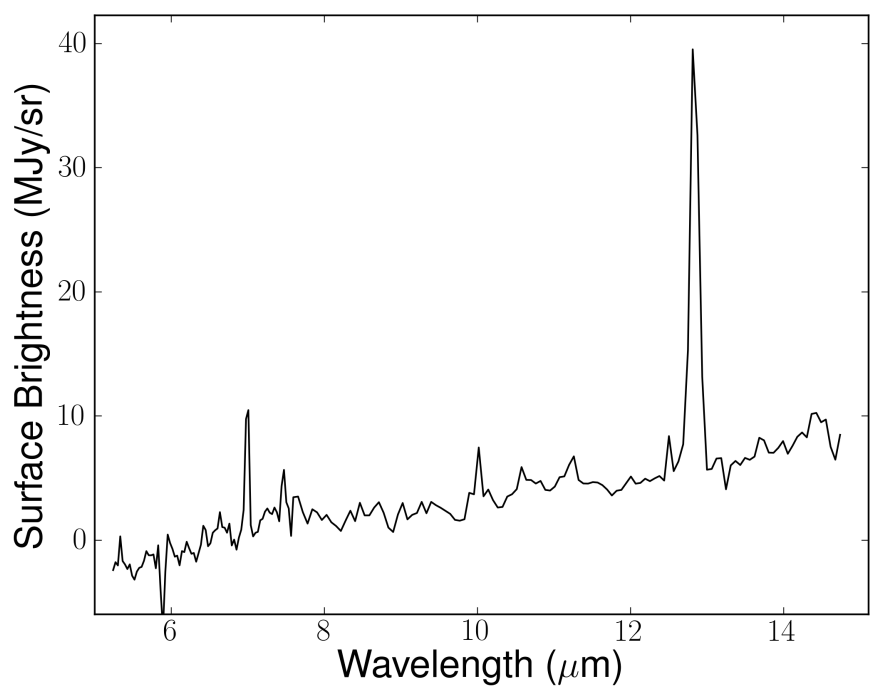}
\caption{Background-subtracted short-low spectrum of the brightest
  emitting radiative knot in N103B. Lines of Ar (7.0 $\mu$m) and Ne
  (12.8 $\mu$m) are present.
\label{shortlow}
}
\end{figure*}

A few arcseconds interior to the forward shock are several knots that
also emit in [S II] and [O III], tracers of regions where the shocked
material has cooled significantly and the shocks have become
radiative. The brightest knot in the remnant at all three optical
wavelengths (approximately halfway in between the north and south
boundaries), which appears unresolved in the image, is also seen in
the IRAC data, also shown in Figure~\ref{images2}. The source of this
emission is not clear, but is likely due to radiative cooling lines in
the near-IR. We show, in Figure~\ref{shortlow}, the short-low spectrum
from this knot, which reveals an Ar II line at 7 $\mu$m. The IRS cuts
off at $\sim 5.5$ $\mu$m, so the source of the emission at IRAC
channels 2 and 3 (4.5 and 5.6 $\mu$m) cannot be identified from the
spectra available.

\begin{figure}
\centering
\includegraphics[width=14cm]{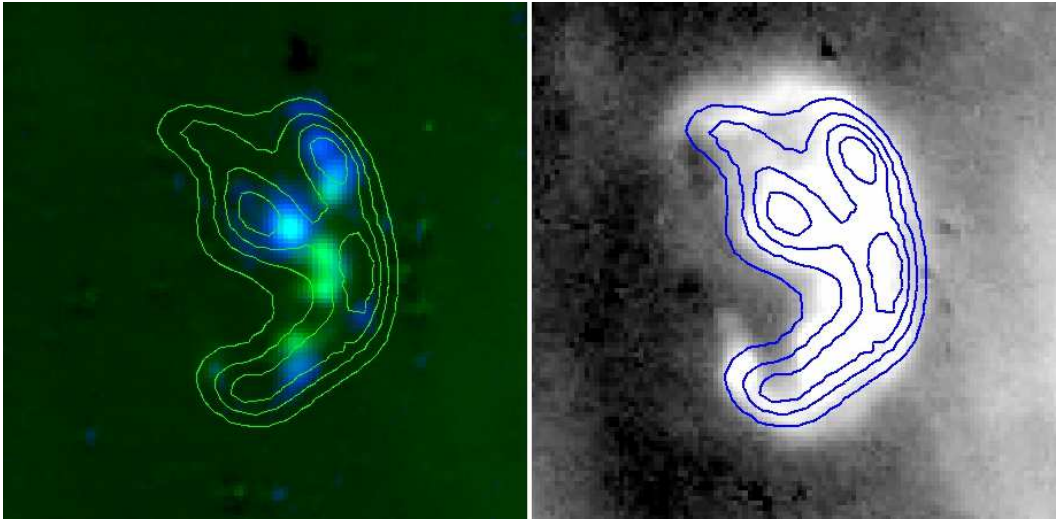}
\caption{{\em Left}: Optical image with [S II] in green and [O III] in
  blue,with contours overlaid from deconvolved MIPS 24 $\mu$m
  image. There is little correlation between dust emission and
  radiative shock emission. {\em Right}: H$\alpha$ emission, stretched
  to show faintest nonradiative emission at forward shock, with the
  same MIPS contours overlaid.
\label{opticaldust}
}
\end{figure}

In Figure~\ref{opticaldust}, we show an optical image in [S II] and [O
  III], highlighting only the portions of the remnant where the shocks
have become radiative. Contours of the 24 $\mu$m emission overlaid on
top of these shocks shows only a weak correlation between the emission
from warm dust and the cooler, radiative shocks. We also show in that
Figure an H$\alpha$ image highlighting the Balmer-dominated portions
of the shock. The correlation with the MIPS 24 $\mu$m emission is much
stronger, consistent with what has been seen in other remnants
resulting from both types of SNe, both young, like Kepler
\citep{blair07}, and old, like the Cygnus Loop \citep{sankrit10}. This
correlation confirms the presence of nonradiative, Balmer-dominated
shocks in N103B. While slow radiative shocks into dense clumps can
heat dust, the much better morphological correlation with the
H$\alpha$ emission means that the nonradiative shocks are the dominant
heating source in this remnant.

\begin{figure*}
\centering
\includegraphics[width=5cm]{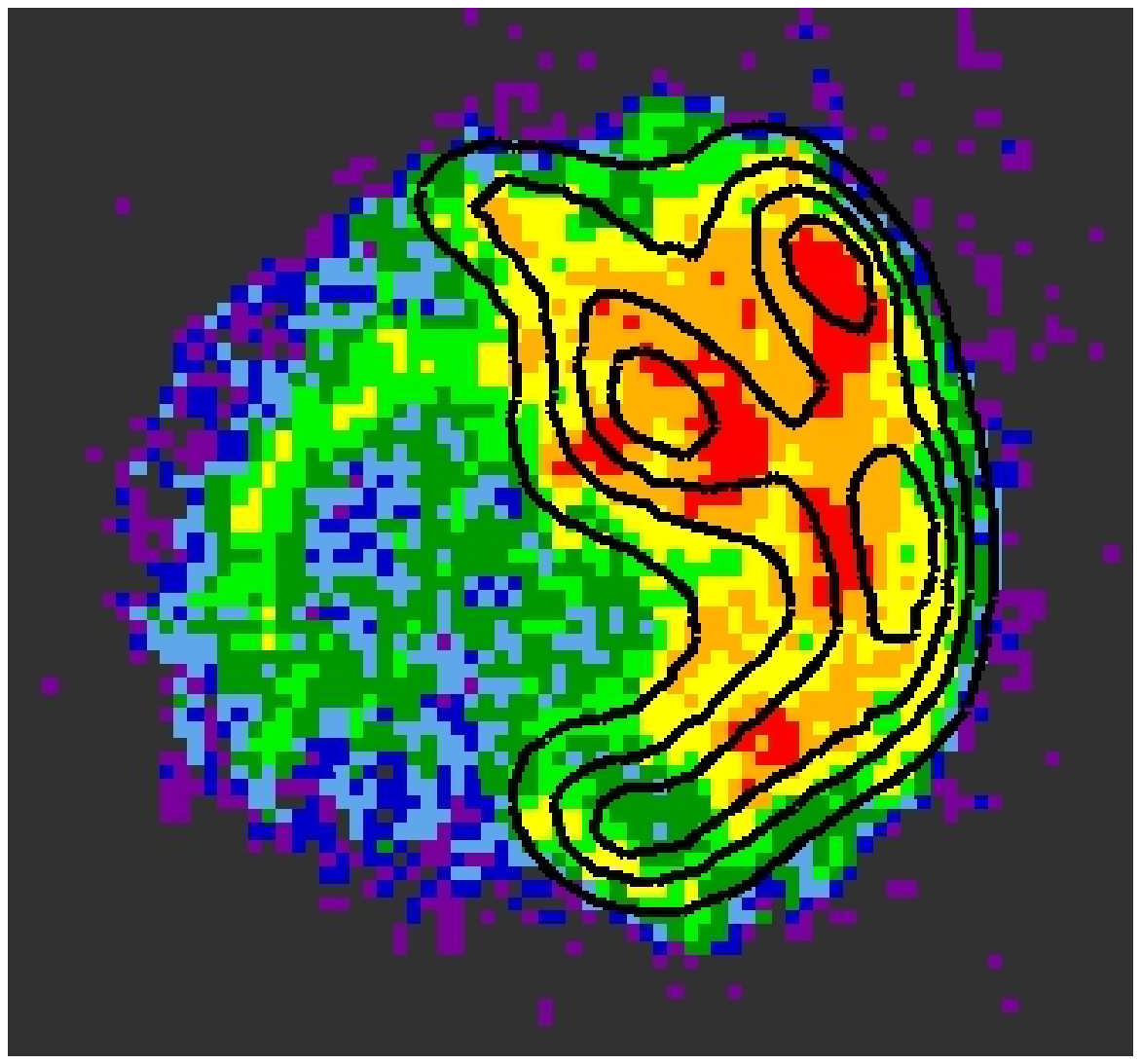}
\includegraphics[width=5cm]{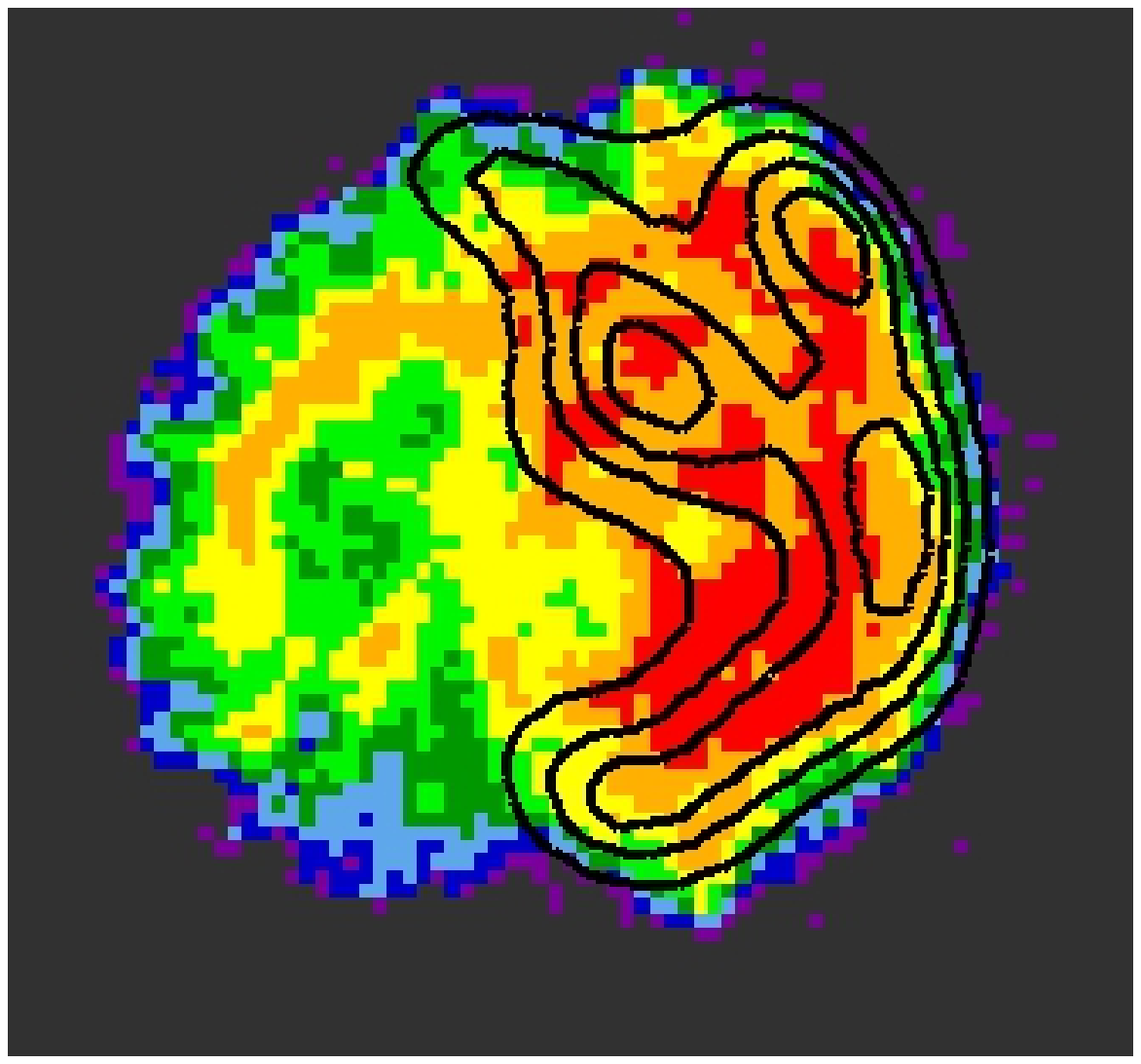}
\includegraphics[width=5cm]{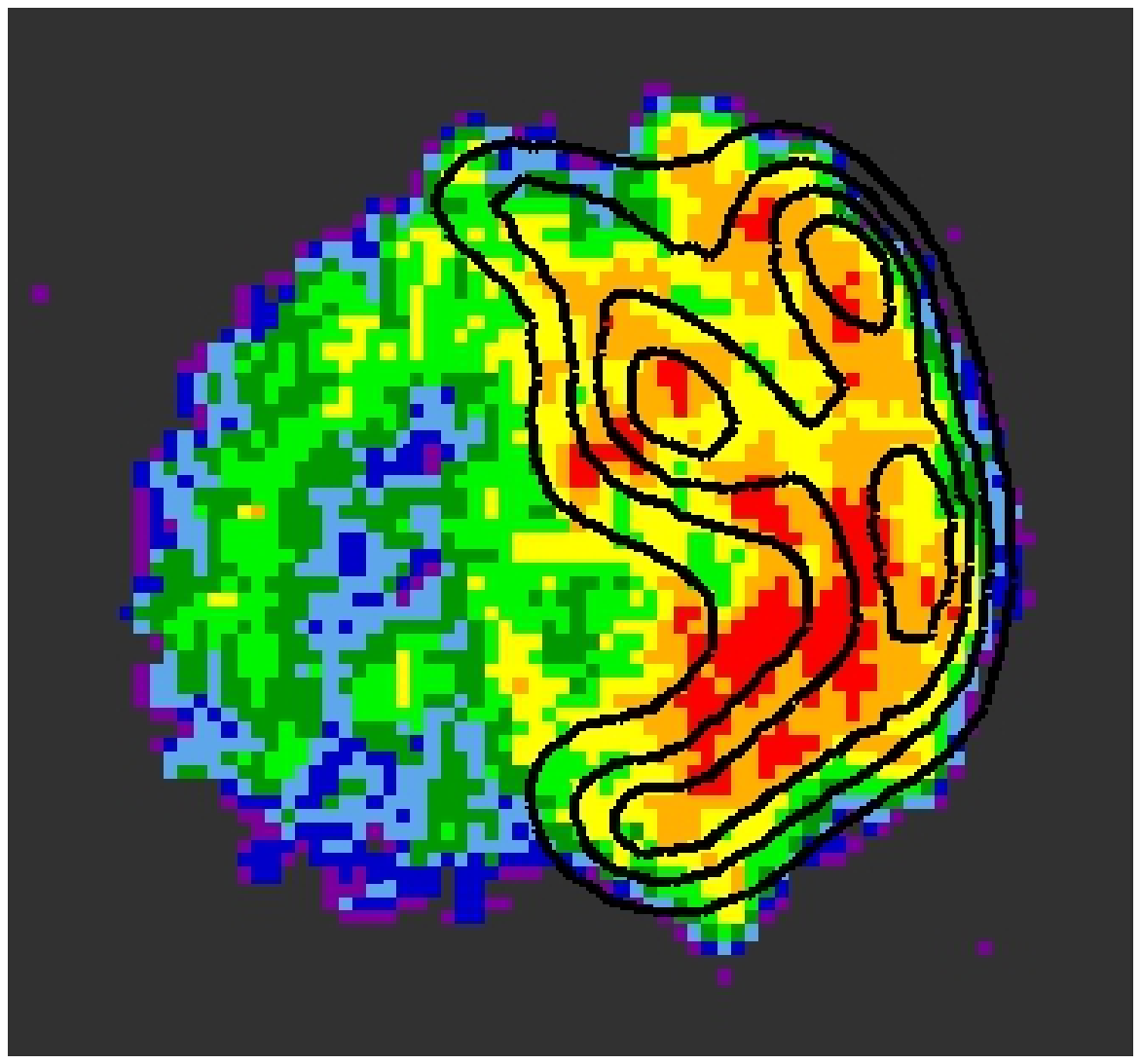}
\caption{{\em Left}: 0.5-0.75 keV X-rays, containing O emission; {\it
    Center}: 0.75-1.2 keV X-rays, containing Fe emission; {\it Right}:
  1.2-6 keV X-rays, containing Si and S emission. Contours from the
  MIPS 24 $\mu$m image are overlaid on each. All images have been
  slightly smoothed with a 1-pixel gaussian.
\label{xraydust}
}
\end{figure*}

{\it Chandra} X-ray images of N103B were first shown by
\citet{lewis03}. The ejecta of SNe Ia are rich in Fe, Si, and S, while
not containing much O. In Figure~\ref{xraydust}, we show narrowband
{\it Chandra} images centered on three different line emission bands:
O, Fe, and Si$+$S, along with contours from the MIPS warm dust
emission. Aside from the emission at all energies being brighter on
the western side of the remnant, where the dust emission is also
strongest, no obvious morphological comparison stands out between the
dust emission and the Fe and Si $+$ S bands. Since no dust has ever
been observed in the ejecta of a Type Ia SN, this is not
surprising. The best correlation with the 24 $\mu$m dust comes from
the 0.5-0.75 keV X-rays, containing significant O emission. Most of
this emission arises from the interaction of the forward shock with
the CSM.

\begin{figure}
\centering
\includegraphics[width=14cm]{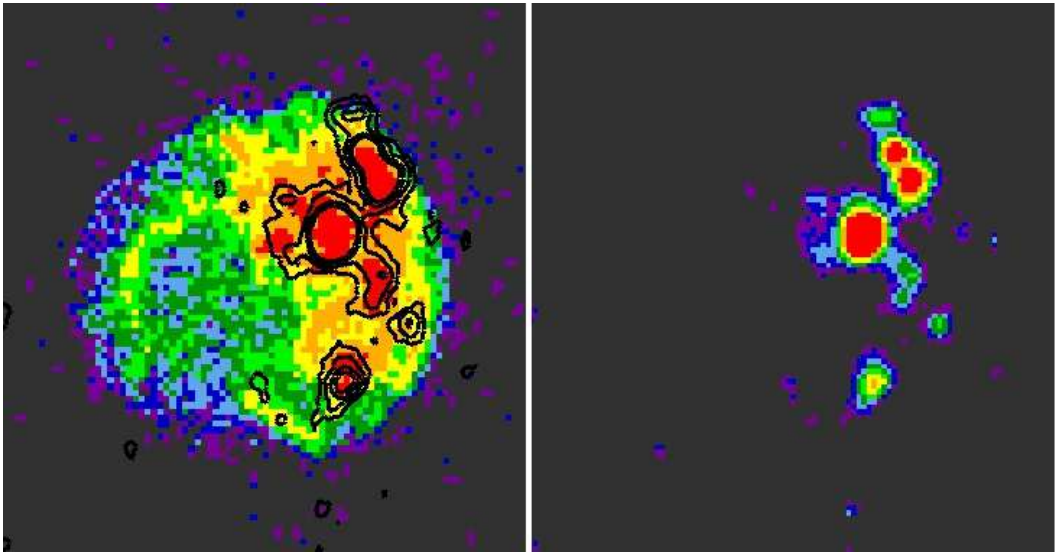}
\caption{{\it Left}: 0.5-0.75 keV X-rays, containing O emission. {\it
    Right}: Optical [O III] image. Contours of [O III] emission are
  overlaid on the X-ray image, highlighting near perfect correlation
  of bright knots in both energy regimes.
\label{oxygenknots}
}
\end{figure}

Interestingly, the brightest X-ray knots of O emission correspond
nearly perfectly to the bright optical [O III] knots, as we show in
Figure~\ref{oxygenknots}. The optically-emitting knots arise from the
densest regions of the CSM, and are where the shocks have become
radiative. The correlation implies that there are multiple shocks
present in these dense clumps; some dense and cool enough to provide
significant optical emission, and some fast and hot enough for X-ray
emission.

\subsection{Asymmetry}

Perhaps the most obvious feature of the remnant at all wavelengths is
the asymmetry between the E and W halves. Stellar winds, by
themselves, cannot easily reproduce such an asymmetry. While winds can
be anisotropic, they should still be symmetric in either the polar or
equatorial direction. Here, again, we are reminded of Kepler, which is
a factor of several brighter in the N hemisphere than the S
\citep{blair07}. Kepler, located well out of the Galactic plane where
densities are $\sim 10^{-2.5} - 10^{-3}$ cm$^{-3}$ \citep{mckee77},
has a high systemic velocity of $\sim 300$ km s$^{-1}$. In the model
of \citet{bandiera87}, the asymmetry of Kepler is reproduced with a
``comet'' model, in which the mass losing star moving at relatively
high velocity creates a bow shock from its interaction with the ISM.

N103B may not have the high spatial velocity of Kepler relative to the
local ISM, but this is not necessary for a bow shock to form. As a
rough calculation, the ram pressure (which scales as $n v^{2}$,
where $n$ is the ISM density and $v$ the velocity of the system)
from the ISM resulting from a $300$ km s$^{-1}$ star moving through an
ISM with $n = 1 \times 10^{-3}$ cm$^{-3}$ (appropriate for Kepler)
is equivalent to the pressure exerted by an ISM with $n = 1$
cm$^{-3}$ on a system moving at 10 km s$^{-1}$. Hydrodynamic
simulations of \citet{villaver12} show that bow shocks with
significant density differences between the direction of motion and
the trailing direction can form in the winds of AGB stars with
velocities as low as 10 km s$^{-1}$ in densities as low as 0.1
cm$^{-3}$. Many such systems have been observed with {\it Herschel};
see \citet{jorissen11} for examples.


\section{Discussion}
\label{disc}

The density we derive is very close to what was found for Kepler's SNR
\citep{williams12}, unsurprising given the similarities between the
two spectra. As we show in Table~\ref{irxtable}, this density is an
order of magnitude or more higher than any other Type Ia
remnants. What could account for such a high density? One possibility
is a coincidental encounter of the shock in the W hemisphere with a
dense interstellar cloud. However, this explanation would not explain
the shape of the remnant as being round and reminiscent of a bow shock
nebula. Additionally, the 18 $\mu$m feature of silicate dust in dense
interstellar clouds is actually at 18 $\mu$m (in fact, the optical
properties of silicate grains which place the location of the feature
at 18 $\mu$m are usually determined from observations of such clouds).

A more straightforward explanation is the same one invoked for Kepler:
N103B is expanding into a dense CSM ejected from the progenitor system
prior to explosion. This would explain the high densities and the
formation of a bow shock nebula. An obvious question that arises from
this is what the progenitor was. In Kepler, an AGB star has been
favored by several authors, but there is an issue with an AGB
interpretation for N103B. The atmospheres of massive AGB stars should
be overabundant in N \citep{marigo11}, particularly at LMC
metallicities \citep{karakas03}, but the optical spectra of N103B do
not show such an N enhancement \citep{russell90}. This can be taken as
evidence that the WD companion had not undergone the first dredge-up
prior to the pre-SN mass loss.  Explanations other than the symbiotic
AGB progenitors have been invoked to account for objects such as SN
2002ic, a Type Ia with the dense CSM.  In the models of \citet{han06}
(see also similar models in \citet{meng10}), material from a companion
star is lost prior to the first dredge-up and the RGB and AGB phases
of the stellar evolution.

Whatever the interpretation of the dense material in N103B, an
additional issue is that with an ionization timescale from a
plane-shock model for the X-ray emitting gas of 10$^{11}$ cm$^{-3}$ s
\citep{lewis03}, our fitted post-shock density leads to a shock age of
$\sim 60$ yr. We expect a shock age of $\sim 300$ yr, since the
``effective'' age in Sedov dynamics for a plane-shock model is about
1/3 the true age of the remnant \citep{borkowski01}. One possibility
is that this ionization timescale is too short; \citet{vanderheyden02}
give a value of 2.3 $\times 10^{11}$ cm$^{-3}$ s. Using this, our
shock age would be 140 yr. Another possibility is that the remnant is
younger than the best-fit age of 870 yr reported in
\citet{rest05}. The authors there give a lower limit of 380 yr, a
value that would further reduce the discrepancy by a factor of
two. While this SN would have been easily visible from Earth with the
naked eye, there is no documented historical record. However, this is
not necessarily a problem, as another SN Ia in the LMC with an age
determined from light echoes, SNR 0509-67.5 (an age of $\sim 400$ yr),
also has no documentation of its SN event. \citet{badenes08} present
an interesting discussion of the lack of a historical record for
0509-67.5 in the context of exploration and colonization of the
southern hemisphere.

Since the remnant is basically unresolved in the spectroscopic
observations, where our 3-D spatial-spectral cube has a spatial
resolution of $>10''$ beyond 30 $\mu$m, another possible resolution to
this is that we are seeing a superposition of various physically
distinct emitting regions. This is the case in Kepler, where the
density varies significantly throughout the remnant
\citep{williams12}. As we showed for Kepler, dust emission at short
wavelengths is dominated by nonradiative shocks driven into denser
than average material, while at longer wavelengths, faster shocks into
less dense material will dominate. Unfortunately, this hypothesis is
not testable for N103B with current data. As we stated in
Section~\ref{observations}, flux measurements at 70 $\mu$m are
unreliable, and no emission at longer wavelengths is seen from {\it
  Herschel}. In the near future, the only reasonable observational
test of this in the IR would be much higher spatial-resolution IR
spectra. This will be possible with the {\it James Webb Space
  Telescope (JWST)}. Even though JWST will only go out to $\sim 25$
$\mu$m, one could still search on arcsecond scales for variations in
the shape of the continuum emission, a clear indication of varying
dust temperatures, and thus, varying densities.

\section{Conclusions}
\label{conclusions}

The LMC remnant N103B shows strong emission in the mid-IR from warm
dust grains in the post-shock environment. Its luminosity at 24 $\mu$m
is nearly an order of magnitude higher than any other known Type Ia
SNR. The spectrum is virtually identical to that seen from Kepler's
SNR, which is known to be interacting with a dense CSM. We propose
that N103B is a more evolved version, i.e., an ``older cousin,'' of
Kepler. We have several lines of circumstantial evidence for a
circumstellar medium. The densities observed are substantially higher
than any other Type Ia remnant known (with the exception of Kepler),
either in the Galaxy or the LMC. The ``18 $\mu$m'' feature of silicate
dust is clearly different from other silicate dust observed in the
ISM, possibly indicating that it has not yet been processed by the
ISM. The shape of the remnant, strongly indicative of a bow-shock
formation in the W, requires an outflow from the progenitor system.

This interpretation is not without issues. The densities observed
imply a very young shock (or a very recent shock interaction), lower
by a factor of several from what is expected. A younger age and/or
lower densities in the nonradiative shocks would alleviate
this. Whereas Kepler has been interpreted as having an AGB star in the
progenitor system, the overabundance of N in the optical spectrum
predicted for massive AGB stars is not observed in N103B. However, the
detailed relationship between the time evolution of N production in
the progenitor system and the time of the SN explosion is poorly
understood. One solution to this is that it may not have been an AGB
star in the progenitor system, but rather a star that experienced
significant mass prior to the first dredge-up of material. Such an
explanation has been invoked to explain the Type Ia SN
2002ic. Finally, it is entirely possible that we are seeing an
unresolved mix of ISM and CSM gas, which would surpress any N
enhancement in the optical spectra.

Alternative scenarios to the CSM interpretation are less
appealing. One possibility to explain the high density observed would
be a coincidental encounter of the shock in the W hemisphere with a
dense interstellar cloud, but such a chance encounter would not
necessarily solve the issues listed above. While this could explain
the lack of N overabundance (since the dense gas would be of ISM
origin, and not CSM), it would not explain the shape of the remnant as
being round and reminiscent of a bow shock nebula. Additionally, the
18 $\mu$m feature of silicate dust in dense interstellar clouds is
actually at 18 $\mu$m (in fact, the optical properties of silicate
grains which place the location of the feature at 18 $\mu$m are
usually determined from observations of such clouds).

If the CSM interpretation is correct, N103B would become only the
second known example of a Type Ia SN to be interacting with such a
medium during the remnant phase. Such dense surroundings for the
forward shock account for the strong O emission seen in the X-ray
spectra. The strong E-W asymmetry is very similar to Kepler (where the
asymmetry is N-S). A slowly moving progenitor system blowing a slow
wind can give rise to such a geometry. Such winds are most easily
produced in a single-degenerate scenario. It is possible, though far
from certain, that a companion star may still exist in the vicinity of
N103B. However, because the lack of N abundance in the spectra
disfavors an AGB or RGB star (both of which are quite bright),
detecting such a companion star at 50 kpc may not be an easy task.

A dense CSM suggests that the SN that created N103B arose from a
binary system containing a relatively young, massive progenitor. In an
extragalactic population study, \citet{aubourg08} report significant
evidence for a small population of short-lived SNe Ia progenitors with
lifetimes less than 180 Myr and main-sequence masses in the range of
$\sim 3.5-8$ $\msun$. \citet{mannucci06} also concluded that a
population of ``prompt'' SNe Ia must exist (characterized by delay
times from stellar birth to SNe of $\sim 100$ Myr). \citet{badenes09}
note that the local environs of N103B are associated with a burst of
star-forming activity, with a prominent extended peak between 100 and
50 Myr in the past. Their statistical analysis gives a 73\% chance
that a Type Ia SN in such a region would be a prompt Ia, and they note
that a metal-rich progenitor is likely. {\it Spitzer's} limited
resolution does not allow for spatially-resolved spectroscopy, so the
potential for future study in the mid-IR with the {\it James Webb
  Space Telescope} is clear. Better characterization of the X-ray
emission, such as spectrally decomposing regions dominated by the
forward-shocked CSM and the reverse-shocked ejecta on small spatial
scales would be useful. A deep {\it Chandra} observation could
accomplish this on scales similar to what will be observable with {\it
  JWST}.

\acknowledgments

\newpage
\clearpage

\begin{deluxetable}{llcccl}
\tablecaption{Imaging Observations of N103B from CTIO 1.5\,m Telescope}
\tablewidth{0pt}
\tablehead{ & \multicolumn{3}{c}{Filter} &  
\colhead{Exposure} & \\
\cline{2-4} 
\colhead{Date} & \colhead{Designation} & \colhead{ $\lambda_0$\ (\AA)} &\colhead{$\Delta\lambda$\tablenotemark{a}\ (\AA)}  & 
\colhead{ (s) }& \colhead{Observer}  
}
\startdata
1994 Dec 7 \phn & \oiii\ \lam5007 &5008  & \phn58 & $2  \times 750 $ & C. Smith \\
1994 Dec 7 \phn &Green Continuum &5133  & 100 &  $2 \times 500 $ & C. Smith \\
1994 Jan 8 \phn &\ha & 6565  &  \phn24 & $2  \times 600 $ & Winkler \\
1994 Dec 7 \phn &\sii\ \lamlam 6716, 6731 & 6728  &  \phn48 &  $2  \times 900 $ & C. Smith\\
1994 Jan 8 \phn &Red Continuum &6848  &  \phn94  & $2  \times 400 $ & Winkler \\
\enddata
\tablenotetext{a}{FWHM}  
\label{optical}
\end{deluxetable}

\newpage
\clearpage

\begin{deluxetable}{lc}
\tablecolumns{3}
\tablewidth{0pc}
\tabletypesize{\footnotesize}
\tablecaption{{\it Spitzer} IR Flux Measurements of N103B}
\tablehead{
\colhead{Wavelength ($\mu$m)} & Flux (Jy) }

\startdata

16 & 0.19 $\pm\ 0.014$\\
24 & 0.48 $\pm\ 0.03$\\
70 & $< 1.3$\\

\enddata \tablecomments{Background-subtracted fluxes measured from a
  circular aperture centered on the remnant, $40''$ in
  diameter. Uncertainties are combined statistical and systematic
  uncertainties; see text for details.}
\label{irflux}
\end{deluxetable} 

\newpage
\clearpage

\begin{deluxetable}{lccc}
\tablecolumns{6}
\tablewidth{0pc}
\tabletypesize{\footnotesize}
\tablecaption{Comparison of IR Emission from Type Ia SNRs}
\tablehead{
\colhead{Object} & n$_{H}$ (cm$^{-3}$) & L$_{\rm IR}$ & L$_{\rm 24, 50 kpc}$ }

\startdata

0548-70.4 & 1.7 & 2.1 $\times 10^{36}$ & 2.63\\
0509-67.5 & 0.59 & 1.5 $\times 10^{36}$ & 16.7\\
Tycho's SNR & 0.4 & 5.4 $\times 10^{36}$ & 73\\
DEM L71 & 2.3 & 1.2 $\times 10^{37}$ & 88\\
0519-69.0 & 6.2 & 4.3 $\times 10^{36}$ & 92\\
Kepler's SNR & 42 & 2.8 $\times 10^{36}$ & 95\\
N103B & 45 & 8.4 $\times 10^{36}$ & 480\\

\enddata \tablecomments{n$_{H}$ is post-shock density. L$_{IR}$ is the
  luminosity, in units of ergs s$^{-1}$. L$_{24, 50 kpc}$ is the flux
  at 24 $\mu$m, normlized to the LMC distance for Kepler and Tycho
  (assuming 5 kpc distance for Kepler and 3 kpc for Tycho.)}
\label{irxtable}
\end{deluxetable}

\newpage
\clearpage

\end{document}